\documentclass[12pt]{iopart}
\usepackage{graphicx}
\begin{document}
\title{Energy fluctuations at the multicritical point in
two-dimensional spin glasses}
\author{Hidetoshi Nishimori, Cyril Falvo and Yukiyasu Ozeki}
\address{Department of Physics, Tokyo Institute of Technology, 
Oh-okayama, Meguro-ku, Tokyo 152-8551, Japan}
\begin{abstract}
We study the two-dimensional $\pm J$ Ising model, three-state Potts 
model and four-state Potts model, by the numerical transfer matrix method 
to investigate the behaviour of the sample-to-sample fluctuations of the 
internal energy on the Nishimori line. The result shows a maximum at the 
multicritical point in all the models we investigated. The large 
sample-to-sample fluctuations of the internal energy as well as the 
existence of a singularity in these fluctuations imply that the bond 
configuration (or, equivalently, the distribution of frustrated plaquettes)
may be experiencing a non-trivial change of its behaviour at the multicritical
point. This observation is consistent with the picture that the phase 
transition at the multicritical point is of geometry-induced nature.
\end{abstract}
\section{Introduction}
The two-dimensional spins glasses, especially the 
$\pm J$ Ising model and the random $q$-state Potts model, have been 
studied both by theoretical and numerical methods. Some exact results 
have been derived \cite{book}, the phase diagram has been drawn \cite{book,
potts} and recently the exact location of the multicritical point 
has been predicted \cite{pc}.
However the characteristics of the multicritical point are not well understood
except that the critical exponents have different values than those along
the ferro-para critical curve \cite{potts,numeric}. The properties of the 
multicritical point are of great interest because this point represents 
one of the simplest non-trivial renormalization group fixed points in random
spin systems \cite{LeDoussal}
and also because we may be able to understand the significance 
of the Nishimori line from the study of the multicritical point.\par
In the present paper we investigate the sample-to-sample fluctuations of the
internal energy near the multicritical point. The thermal average of the
Hamiltonian, namely the internal energy, has different values from one bond
configuration to another, given a fixed probability parameter ($p$ for the 
ferromagnetic bonds in the $\pm J$ model, for example). Although the internal
energy is self-averaging, the variance of its distribution is of order $N$, 
the system size. If the variance is anomalously large and 
singular at some point, it would 
mean that a non-trivial (and possibly drastic) change happens in the bond 
configuration at that point because the value of the internal energy 
fluctuates strongly from sample to sample. This is exactly what we shall 
show in the present paper to exist at the multicritical point by observing 
the system along the Nishimori line.\par
The Nishimori line plays an important role in the determination of the phase 
diagram. In particular it is believed that the multicritical point lies on the 
line \cite{book}. We have another advantage to study the sample-to-sample 
fluctuations along the Nishimori line: These fluctuations are directly 
related to the average specific heat on the line. This fact reduces numerical 
efforts considerably because it is not necessary to generate very many 
samples to calculate the variance. Only the average of the specific heat 
needs to be computed. We shall make full use of this relation later.\par
In the next section, we show that the sample-to-sample fluctuations of 
the internal energy are very simply related to the specific heat on the Nishimori line.
This facilitates numerical calculations enormously as mentioned above.
In section 3 we investigate numerically 
the behaviour of these fluctuations near the multicritical point
on the Nishimori line for three models: the $\pm J$ Ising model, three-state 
Potts model and four-state Potts model. We will see that these fluctuations show 
a maximum at the multicritical point.
The physical significance of the results is discussed in the final section.
\section{Average fluctuations of the internal energy}
Let us consider the random $Z_q$ model with gauge symmetry which includes 
the $\pm J$ Ising model and the random $q$-state Potts model. The 
general Hamiltonian is written as follows:
\begin{equation}
H = - \sum_{\langle ij \rangle} V(S_i - S_j + J_{ij}),
\end{equation}
where $S_i = 0,1,...,q-1$ is the $q$-state spin variable, $J_{ij} = 0,1,
...,q-1$ is the quenched random bond variable and  $V$ is an interaction function of 
period $q$. The summation is taken over the first-neighbouring sites of the square
lattice.
The $J_{ij}$ variable follows the probability distribution 
\begin{equation}
P(J_{ij};\beta_p) = \frac{\rme^{\beta_pV(J_{ij})}}
{\displaystyle \sum_{\eta=0}^{q-1} \rme^{\beta_pV(\eta)}}.
\end{equation}
The Hamiltonian (1) is invariant under the gauge transformation
$S_i \rightarrow S_i-\sigma_i, J_{ij} \rightarrow J_{ij}+\sigma_i - \sigma_j$,
where $\sigma_i=0,1,...,q-1$,
but $P(J_{ij};\beta_p)$ is not. 
Using the gauge theory \cite{book} some exact results can be derived such as 
the average internal energy on the so-called Nishimori line defined by 
$\beta \equiv 1/T = \beta_p$,
\begin{equation}
[E] \equiv [\langle H \rangle ] = -N_B \frac{\sum_{\eta}V(\eta)
\rme^{\beta V(\eta)}}{\sum_{\eta}\rme^{\beta V(\eta)}}.
\end{equation}
Here the square brackets denote the configurational average, the angular brackets
are for the thermal average and $N_B$ is the number of bonds.
If we denote
\begin{equation*}
\{V\} = \sum_{\eta=0}^{q-1}V(\eta)P(\eta;\beta),
\end{equation*}
equation (3) is expressed as
\begin{equation}
[E] = -N_B\{V\}.
\end{equation}
In the case of the $\pm J$ Ising model, $q=2$, $V(0)=+J$ and $V(1) = -J$,
the average internal energy (3) reduces to
\begin{equation*}
[E] = -N_BJ\tanh(\beta J). 
\end{equation*}
In the case of the $q$-state Potts model with $V(0) = +J$ and $V(1)=V(2)=...=
V(q-1)=0$,
\begin{equation*}
[E] = -N_BJ\frac{1}{1+(q-1)\rme^{-\beta J}}.
\end{equation*}\par
We also find from the gauge theory that
\begin{equation}
[\langle H^2 \rangle ]=N_B\{V^2\} + N_B(N_B-1)\{V\}^2,
\end{equation}
when $\beta=\beta_p$, thus, with equation (4),
\begin{equation}
N_B(\{V^2\}-\{V\}^2) = [\langle H^2 \rangle] - [\langle H \rangle ]^2.
\end{equation}
Since the specific heat is defined by
\begin{equation}
T^2[C] = [ \langle H^2 \rangle - {\langle H \rangle}^2],
\end{equation}
we obtain, combining equations (6) and (7), 
\begin{equation}
N_B(\{V^2\}-\{V\}^2)-T^2[C] = [E^2] - [E]^2 ( \ge 0 ).
\end{equation}
The right-hand side of equation (8) is the sample-to-sample 
fluctuations of the internal energy, $[(\Delta E)^2]$. Note that equation (8) 
holds only on the Nishimori line $\beta=\beta_p$.
In the case of the $\pm J$ Ising model equation (8) reduces to the 
well-established inequality
\begin{equation*}
T^2[C] \le N_BJ^2 {\rm sech}^2 (\beta J).
\end{equation*}\par
Equation (8) indicates that the sample-to-sample fluctuations of the
internal energy can be evaluated from the average of the specific heat
because the quantity $\{V^2\}-\{V\}^2$ is trivially calculated.
This fact is a great advantage since it eliminates the necessity to evaluate 
the variance directly  by using a large number of samples; the specific heat
is a self-averaging quantity and therefore can be calculated with high 
precision from a small number of samples with sufficiently large system size.
Away from the Nishimori line, we do not have such a simple relation because
equations (4) and (5) do not hold.\par
\begin{figure}
\begin{center}
\includegraphics[width=0.25\linewidth]{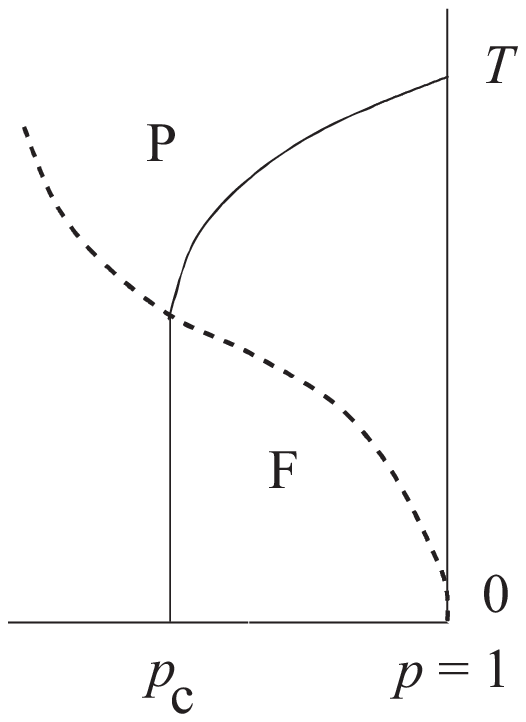}
\end{center}
\caption{\label{diagram} Phase diagram of the $\pm J$ Ising model in two
dimensions. There exist ferromagnetic (F) and paramagnetic (P) phases,
and no spin glass phase exists. The Nishimori line is shown dashed.}
\end{figure}

Since the actual distribution of bond values (and therefore of frustration)
changes from sample to sample in finite-size systems with given $\beta_p$,
the internal energy takes various values from one set of bonds to another
around the average value. The relation (8) helps us to estimate the variance 
of the internal energy $[(\Delta E)^2]$, and directly depends on the 
distribution of frustration, which has a geometrical nature: 
The sample-to-sample fluctuations of the internal energy should 
reflect the geometrical structure of the distribution of frustration,
which is a quantity independent of thermal variables but depends only
on the bond configuration. This argument suggests that we may be 
able to investigate the properties of the distribution of frustration
by looking at the sample-to-sample fluctuations of the internal energy,
or equivalently, the average specific heat on the Nishimori line.
It is believed that the Nishimori line intersects the phase boundary 
between paramagnetic and ferromagnetic phases \cite{book} 
(see figure \ref{diagram}). If the distribution of frustration goes under 
a sharp change of its behaviour, it would be reflected in anomalously
large sample-to-sample fluctuations of physical quantities including
the internal energy. Such an anomaly, if it exists on the right-hand
side of equation (8), is related to the singularity of the specific heat
at the multicritical point on the Nishimori line due to equation (8).
Since the specific heat is singular at the multicritical point, we may
conclude that the sample-to-sample fluctuations are also singular there.
Hence, by calculating the average specific heat near the multicritical point,
we are able to investigate the structure of the distribution of frustration,
which is of purely geometrical nature. A similar argument has been used to
relate the free energy and  the entropy of frustration \cite{geometry}. 
The next section gives us some numerical supports to confirm this idea.
\section{Numerical calculations}
We have tested the ideas of the previous section by numerically calculating
the average specific heat on long strips by transfer matrix method.
We have studied three models: the $\pm J$ Ising model, three-state Potts model and 
four-state Potts model. We have used long strips of length 5000 and of various 
widths with free boundary conditions and without external field. 
By the transfer matrix method we calculated 
the free energy along the Nishimori line. The specific heat was obtained from
numerical second derivative of the free energy by the inverse of the temperature.
We averaged the specific heat over ten samples.\par
For the $\pm J$ Ising model the widths of the lattice are  
$3 \le L \le 14$. The lattice size (up to $5000\times 14$) and number of samples (ten)
are much smaller than the existing numerical transfer matrix studies \cite{numeric,Merz}.
However, since our purpose is much more modest than these investigations (not trying 
to evaluate the critical exponents, for instance), our system size turns out to be
sufficient as we shall show. Another reason for the small scale of our computation
is that the use of the formula (8) greatly reduces the necessary number of samples
as discussed in the previous section.\par
We define the parameter $p$ for the $\pm J$ Ising model by
\begin{equation*}
\rme^{2\beta_pJ}=\frac{p}{1-p}.
\end{equation*}
Thus $p$ is the probability that $J_{ij}=0$ $i.e.$ $V(J_{ij}) = +J$ 
\footnote{Notice that we use the Hamiltonian (1) where $J_{ij}$ takes 
the values 0 and 1. This notation is completely equivalent to the usual 
Hamiltonian $H= - \sum J_{ij}S_iS_j$ where $S_i$,$S_j$ and $J_{ij}/J$ 
take the values $\pm 1$.}. Each bond follows the probability distribution
\begin{equation*}
P(J_{ij}) = p\delta(J_{ij}) + (1-p)\delta(J_{ij}-1),
\end{equation*}
and we restrict our studies to $0.5 < p < 1$. For the three-state Potts model, 
$3 \le L \le 9$ and for the four-state Potts model, $3 \le L \le 7$. For the
$q$-state Potts model, we define 
\begin{equation*}
\rme^{\beta_pJ}=\frac{1}{p}-(q-1),
\end{equation*}
and the probability distribution (2) is then
\begin{equation*}
P(J_{ij}) = (1-(q-1)p)\delta(J_{ij}) + \sum_{l=1}^{q-1} p\delta(J_{ij}-l),
\end{equation*}
with $0<p<1/(q-1)$.\par
\begin{figure}
\begin{center}
\includegraphics[width=0.5\linewidth,angle=-90]{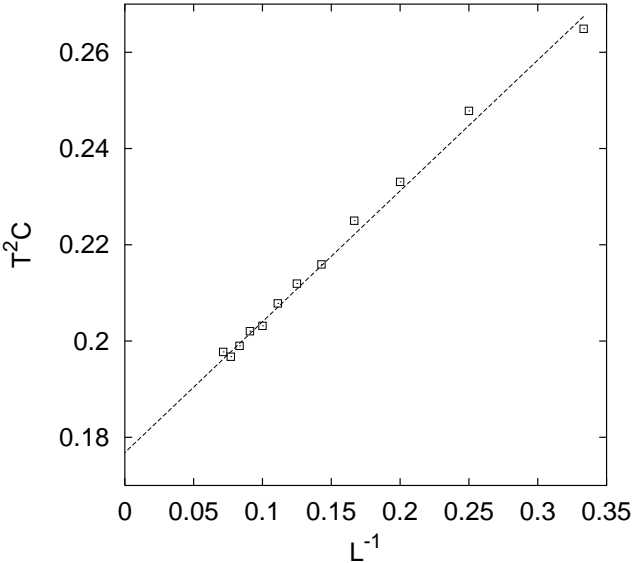}
\end{center}
\caption{\label{pmj1}Extrapolation of $T^2C$ near the multicritical point for the 
$\pm J$ Ising model, $p=0.8900$.} 
\end{figure}
\begin{figure}
\begin{center}
\includegraphics[width=0.5\linewidth,angle=-90]{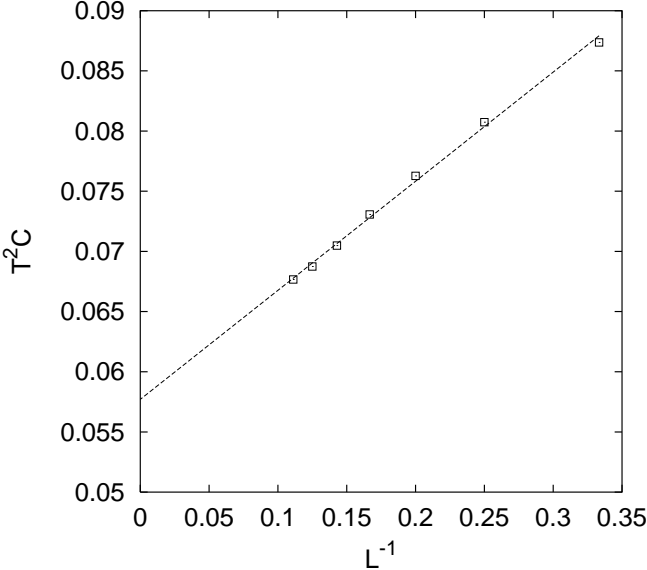}
\end{center}
\caption{\label{potts31}Extrapolation of $T^2C$ near the multicritical point
for the three-state Potts model, $p=0.0795$.}
\end{figure}
\begin{figure}
\begin{center}
\includegraphics[width=0.5\linewidth,,angle=-90]{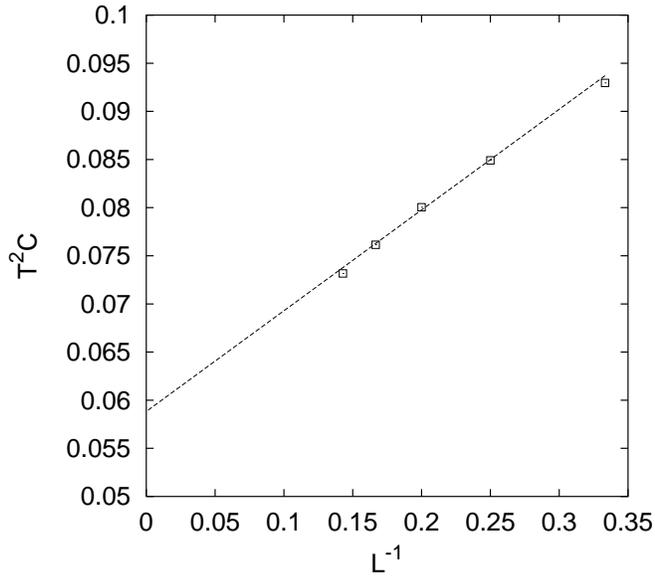}
\end{center}
\caption{\label{potts41}Extrapolation of $T^2C$ near the multicritical point
for the four-state Potts model, $p=0.0615$.}
\end{figure}
\begin{figure}
\begin{center}
\includegraphics[width=0.5\linewidth,angle=-90]{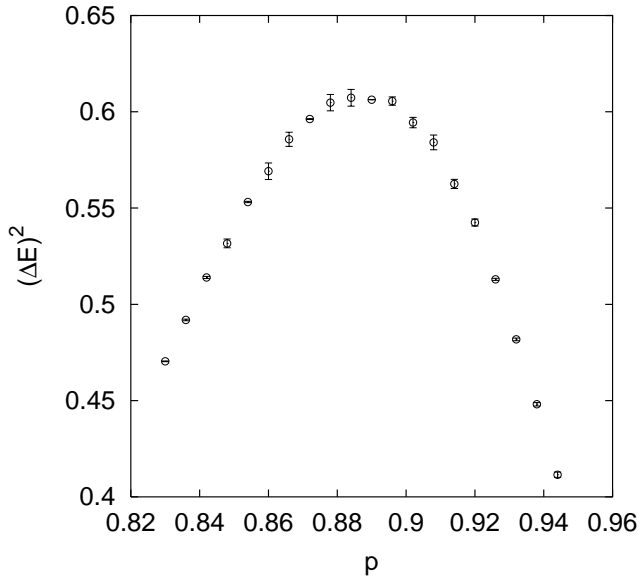}
\end{center}
\caption{\label{pmj2}  Sample-to-sample fluctuations  of the internal energy
for the $\pm J$ Ising model.}
\end{figure}
\begin{figure}
\begin{center}
\includegraphics[width=0.5\linewidth,angle=-90]{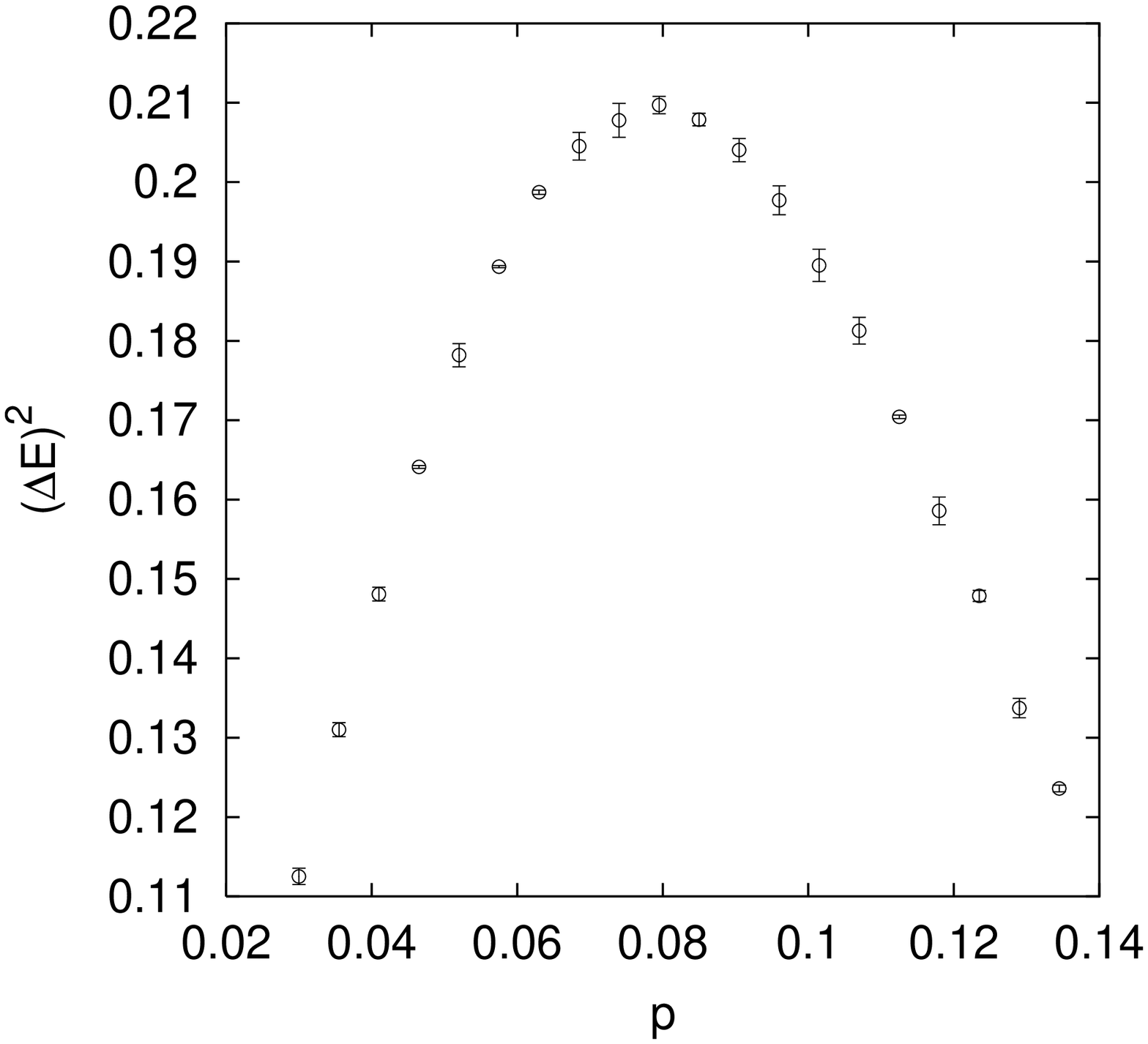}
\end{center}
\caption{\label{potts32} Sample-to-sample fluctuations  of the internal 
energy for the three-state Potts model.}
\end{figure}
\begin{figure}
\begin{center}
\includegraphics[width=0.5\linewidth,angle=-90]{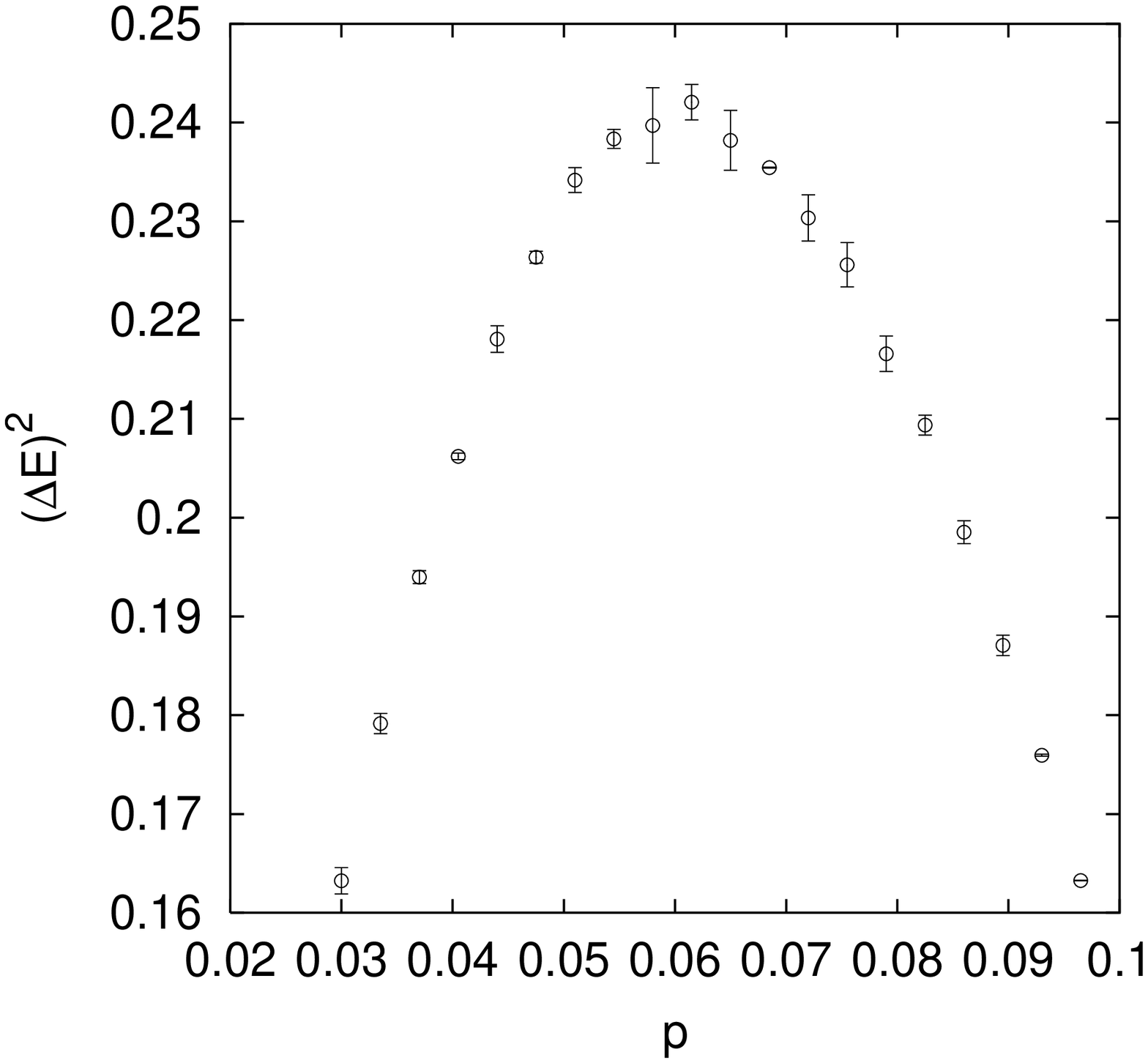}
\end{center}
\caption{\label{potts42} Sample-to-sample fluctuations  of the internal 
energy for the four-state Potts model.}
\end{figure}
For each point on the Nishimori line $\beta=\beta_p$  we extrapolate the 
specific heat to the limit $L \rightarrow \infty $. We have found that
the extrapolation by the $L^{-1}$-law works quite well in almost all cases
(see figures \ref{pmj1}, \ref{potts31} and \ref{potts41}). This is natural
because we are mostly treating values of $p$ away from the multicritical 
point. Exactly at the multicritical point, the extrapolation needs some
care, in particular if one wishes to estimate the critical exponents 
\cite{numeric}. However, for our purpose to confirm the existence of a 
peak in the sample-to-sample fluctuations of the internal energy at the multicritical
point, the method of extrapolation is rather irrelevant, which we confirmed
by using other powers such as $L^{-1.5}$ in some cases.
The error bars in the extrapolated values have been estimated by changing
the range of extrapolation fitting, for example, from comparison of extrapolation
using $3 \le L \le 14$ with the one using $6 \le L \le 14$.\par
The resulting values of the energy fluctuations are plotted in figures 
\ref{pmj2}, \ref{potts32} and \ref{potts42} for the $\pm J$ Ising model,
three-state Potts model and four-state Potts model, respectively.
The position of the multicritical points are 
predicted in reference \cite{pc}: for the $\pm J$ Ising model $p_c=0.889972$, for 
the three-state Potts model $p_c=0.0797308$ and for the four-state Potts model 
$p_c=0.0630965$. By comparison of those values with figures 
\ref{pmj2}, \ref{potts32} and \ref{potts42} we see 
clearly that the sample-to-sample fluctuations of the internal 
energy have a maximum at (or at least very near) the multicritical point 
for each model. It seems reasonable to expect similar behaviour in other
random models with $Z_q$ symmetry.
\section{Conclusion}
We have studied the $\pm J$ Ising model, three-state Potts model 
and four-state Potts model by the numerical transfer matrix method.
We have first shown that the sample-to-sample fluctuations of the 
internal energy are, on the 
Nishimori line, related to the specific heat and are expected to have
a peculiar behaviour at the multicritical point. It was found
indeed from numerical calculations that the sample-to-sample
fluctuations become anomalously large at or very near the multicritical
point in all the models we investigated. Since these fluctuations are related
to the specific heat by equation (8), they should have a singularity at the
multicritical point (as the specific heat is singular there) although it
is difficult to see this effect within our numerical precision. The
existence of a peak at the multicritical point in conjunction with
the existence of a singularity, anyway, suggests
that the bond distribution and the distribution of frustration go under
an anomalous change of the behaviour at $p=p_c$. This point of view is 
in agreement with the argument in reference \cite{geometry} that the phase transition
at the multicritical point along the Nishimori line is induced by geometrical
anomaly \footnote{This never implies that the critical exponents should coincide
with those of the simple conventional percolation transition.}. Since the rate
of the thermal excitations ($\rme^{-\beta J}/ \rme^{\beta J}$) coincides
with the ratio of the bond configurations ($(1-p)/p$) on the Nishimori line,
the physical situations are very special there, which may be one of the reasons
for the existence of a maximum in the configurational variance of the thermal
quantity at the multicritical point along the Nishimori line. It is an interesting
future problem to investigate what happens on the sample-to-sample fluctuations
of physical quantities away from the Nishimori line.

{\em Note added in proof.}
We have found after submission of the paper that the sample-to-sample fluctuations
of the energy, equation (8), is proportional to the derivative of the energy by $K_p$
with $K$ fixed.  Thus a maximum in these fluctuations means that the energy changes
most dramatically at the multicritical point as $K_p$ changes.
The exact upper bound on the specific heat is the sum of two derivatives by
$K$ and $K_p$ and corresponds to the derivative along the Nishimori line. 

\end{document}